\documentstyle[12pt,titlepage]{article}
\input epsfig.sty

\font\grande=cmr9.5 scaled \magstep4
\font\medio=cmr9.5 scaled \magstep2
\outer\def\beginsection#1\par{\medbreak\bigskip
      \message{#1}\leftline{\bf#1}\nobreak\medskip
\vskip-\parskip
      \noindent}

\def\laq{\raise 0.4ex\hbox{$<$}\kern -0.8em\lower 0.62
ex\hbox{$\sim$}}
\def\gaq{\raise 0.4ex\hbox{$>$}\kern -0.7em\lower 0.62
ex\hbox{$\sim$}}

\begin{document}
\bibliographystyle {unsrt}

\titlepage

\begin{flushright}
CERN-PH-TH/2006-180
\end{flushright}

\vspace{15mm}
\begin{center}
{\grande Kink-antikink, trapping bags }\\
\vspace{5mm}
{\grande and five-dimensional Gauss-Bonnet gravity}\\
\vspace{15mm}
 Massimo Giovannini 
 \footnote{Electronic address: massimo.giovannini@cern.ch} \\
\vspace{6mm}

\vspace{0.3cm}
{{\sl Centro ``Enrico Fermi", Compendio del Viminale, Via 
Panisperna 89/A, 00184 Rome, Italy}}\\
\vspace{0.3cm}
{{\sl Department of Physics, Theory Division, CERN, 1211 Geneva 23, Switzerland}}
\vspace*{2cm}

\end{center}

\vskip 2cm
\centerline{\medio  Abstract}
Five-dimensional Gauss-Bonnet gravity, with one 
warped extra-dimension, allows classes of solutions where 
two scalar fields combine either in a kink-antikink system or in a 
trapping bag configuration. While the kink-antikink system 
can be interpreted as a pair of gravitating domain walls with 
opposite topological charges, the trapping bag solution 
consists of a domain wall supplemented by a non-topological defect.  In both classes of solutions, for large absolute values of the bulk coordinate (i.e. far from the core of the defects), 
the geometry is given by five-dimensional anti-de Sitter space.
\noindent

\vspace{5mm}

\vfill
\newpage
It is  known since the pioneering works of Lanczos \cite{lan} (see also \cite{lov}) that, in more than four space-time dimensions,
the Einstein-Hilbert action can be supplemented by 
the so-called Euler-Gauss-Bonnet combination (see also \cite{mad2}  for a recent review). Such an inclusion 
leads to field equations that involve, at most, second derivatives 
of the metric.  The Gauss-Bonnet combination arises 
also naturally as first correction in the string tension expansion to the 
low-energy string effective action \cite{mad,zw,des, ts,cal,s}. 

An apparently unrelated observation is that, in the 
presence of infinite extra-dimensions, fields of various spin 
may be localized around higher dimensional defects 
(see, for instance, \cite{rub}). 
Indeed, in the past few years, various analytical solutions containing 
gravitating defects have been discussed either in the context 
of Einstein-Hilbert gravity or in the framework of Brans-Dicke gravity 
\cite{KT,GR,MG,KK,RM,FP}.
Some of these solutions are compatible with five-dimensional anti-de Sitter 
space-time (in what follows ${\mathrm AdS}_{5}$) for 
large absolute value of the bulk coordinate, providing, in this way 
 a smooth realization of the Randall-Sundrum set-up 
where the matter content is given by branes (i.e. gravitating kinks)
of finite thickness.

The purpose of the present paper is to show that, in Gauss-Bonnet gravity,  there exists solutions compatible with ${\mathrm AdS}_{5}$ and containing 
{\em pairs} of gravitating defects rather than a single defect.  Solutions have been obtained in the presence of Gauss-Bonnet gravity but only in the case 
of single defects \cite{MG2,COR,GER,NICK}.

Pairs of defects are known to exist in $(1+1)$ field theories in flat space-time and in the presence of appropriately non-linear interaction potentials 
\cite{RAJA,RAJA2, MONT}. As a consequence of the intrinsic non-linearity 
of the problem, exact solutions are rare even if specific methods have been 
devised in order to deal with the integration of the systems in rather general terms(see \cite{RAJA2,SOU} and references therein).
In the presence of gravity it is more difficult to reduce the problem to the 
quadrature and to find analytical solutions. This difficulty 
 is even more severe in Gauss-Bonnet gravity.

Consider then the case
where the gravity part of the action takes the form\footnote{
The signature of the metric 
is mostly minus, i.e. $(+,-,-,-,-,-)$. Latin (capital) 
indices run over the five-dimensional space-time; Greek indices run over the $(3+1)$-dimensional space-time with Minkowskian signature.}
\begin{equation}
S_{\rm g} =- \int d^{5} x \sqrt{|G|} \biggl(  \frac{R}{ 2 \kappa} + \alpha' {\mathcal R}_{\rm EGB}^2\biggr), 
\label{actiong}
\end{equation}
where $G_{AB}$ is the metric tensor, $R$ is the Ricci scalar and
\begin{equation}
{\mathcal R}_{\rm EGB}^2 = R^2 - 4 R_{AB}R^{AB} + R_{ABCD}R^{ABCD},
\label{EGB}
\end{equation}
is the Euler-Gauss-Bonnet (EGB) combination. In Eq. (\ref{actiong}),
$\kappa = 8\pi G_{5} = 8\pi/M^3$ and $\alpha'$ has dimensions of an 
energy scale, i.e., in natural units, an inverse length.

The matter part of the action includes two scalar degrees of freedom, denoted 
by $\phi$ and $\chi$, interacting via the potential $W(\phi,\chi)$:
\begin{equation}
S_{\rm m} = \int d^{5} x \sqrt{|G|}\biggl[ \frac{1}{2} G^{AB}\partial_{A} \phi
\partial_{B} \phi +\frac{1}{2} G^{AB}\partial_{A} \chi
\partial_{B} \chi - W(\phi,\chi) \biggr], 
\label{actionm}
\end{equation}
The total action will then be given by the sum of the gravity and matter 
action, i.e. $S_{\rm t} = S_{\rm g} + S_{\rm m}$.
The equations of motion are obtained by taking the functional 
derivative of $S_{\rm t}$ with respect to the metric tensor and with respect to the two scalar fields. Functional derivation with respect to the metric tensor leads to the generalized Einstein-Lanczos equations 
\begin{equation}
R_{A}^{B} - \frac{1}{2} \delta_{A}^{B} = \kappa {\mathcal T}_{A}^{B} 
- 2\alpha' \kappa {\mathcal Q}_{A}^{B},
\label{EL}
\end{equation}
where 
\begin{eqnarray}
&&{\cal T}_{A}^{B} = \partial_{A}\phi \partial^B\phi + \partial_{A}\chi \partial^{B} \chi - 
\delta_{A}^{B} \biggl[ \frac{1}{2} G^{M N} \partial_{M} \phi \partial_{N} \phi + 
\frac{1}{2} G^{M N} \partial_{M} \chi \partial_{N} \chi - W(\phi,\chi)\biggr],
\label{e1a}\\
&& {\cal Q}_{A}^{B} = \frac{1}{2} \,\delta_{A}^{B}\, {\cal R}^2_{\rm EGB} -
2\,R\, R_{A}^{B} + 4\,R_{A\,C} \, R^{C\,B} + 4\, R_{CD}\,R_{A}^{~~C\,B\,D}
-2 \,R_{A\,C\,D\,E}\, R^{B\,C\,D\,E}, 
\label{e1b}
\end{eqnarray}
are, respectively, the energy-momentum tensor and  the Lanczos tensor. Functional derivation with respect to $\phi$ and $\chi$ produces 
the following pair of Klein-Gordon equations:
\begin{equation}
G^{A B} \nabla_{A} \nabla_{B} \phi + \frac{\partial W}{\partial \phi} =0,
\qquad G^{AB} \nabla_{A} \nabla_{B} \chi + \frac{\partial W}{\partial \chi} =0,
\label{kg}
\end{equation}
where, we recall, $\nabla_{A} \nabla_{B} = \partial_{A} \partial_{B} - \Gamma_{A B}^{C}\partial_{C}$ when applied to a scalar degree of freedom.

In the case of a five-dimensional warped metric of the type characterized by a bulk coordinate $w$, i.e.
\begin{equation}
ds^2 = a^2(w) [ \eta_{\mu\nu} dx^{\mu} dx^{\nu} - dw^2],
\label{lineel}
\end{equation}
Denoting with the prime a derivation with respect to $w$, the explicit 
form of Eq. (\ref{EL}) becomes:
\begin{eqnarray}
&& {\mathcal H}' \biggl( 1 - \frac{2 \epsilon {\mathcal H}^2}{a^2}\biggr) 
+ {\mathcal H}^2 \biggl( 1 - \frac{2 \epsilon {\mathcal H}'}{a^2}\biggr) = - \frac{\kappa}{3} \biggl[ \frac{{\phi'}^2}{2} + \frac{{\chi'}^2}{2} + a^2 W(\phi,\chi)\biggr],
\label{EL1}\\
&& {\mathcal H}^2 \biggl( 1 - \frac{2 \epsilon {\mathcal H}^2}{a^2}\biggr) 
= \frac{\kappa}{6} \biggl[ \frac{{\phi'}^2}{2} + \frac{{\chi'}^2}{2} - a^2 W(\phi,\chi)
\biggr],
\label{EL2}
\end{eqnarray}
where ${\mathcal H} = (\ln{a})'$. In Eqs. (\ref{EL1})--(\ref{EL2})
 the quantity $\epsilon= 2 \kappa \alpha'$ 
has been also defined and it has dimensions, in natural units, of a length squared.
Using Eq. (\ref{lineel}) into Eqs. (\ref{kg}) the following explicit equations are obtained:
\begin{equation}
\phi'' + 3 {\mathcal H} \phi' - a^2 \frac{\partial W}{\partial \phi} =0,
\qquad \chi'' + 3 {\mathcal H} \chi' - a^2 \frac{\partial W}{\partial \chi} =0. 
\label{kgex}
\end{equation}
By combining Eqs.  (\ref{EL1}) and (\ref{EL2}), the  explicit from of the Einstein-Lanczos equations can be also written as 
\begin{eqnarray}
&&( {\phi'}^2 + {\chi'}^2)= \frac{3}{\kappa} ({\mathcal H}^2 - {\mathcal H}') \biggl[ 1 - \frac{4 \epsilon}{a^2} {\mathcal H}^2\biggr],
\label{EL1a}\\
&& W = - \frac{3}{2 a^2\,\kappa}\biggl\{ ({\cal H}^2 + {\cal H}') \biggl[ 1 - \frac{4\epsilon}{a^2} {\mathcal H}^2\biggr] + 2 {\mathcal H}^2 \biggr\}.
\label{EL2a}
\end{eqnarray}

Consider then the situation where the warp factor 
 tends to ${\mathrm AdS_{5}}$ 
for large absolute value of the bulk coordinate $w$.
A possible choice of warp factor with the desired properties is
\begin{equation}
a(w) = \frac{a_0}{\sqrt{b^2 w^2 + 1}},\qquad {\mathcal H}= -\frac{b^2 w}{b^2 w^2 + 1},
\qquad {\cal H}' =\frac{b^2(b^2 w^2 -1)}{(b^2 w^2 + 1)^2},
\label{geometry}
\end{equation}
where $a_0$ is a free parameter that will be determined 
from the compatibility with the whole system of equations.
In Eq. (\ref{geometry}) the first relation is the ansatz for the warp factor 
while the remaining relations follow from the definition of ${\mathcal H}$ in terms of $a(w)$.

The method employed in order to find the solution is constructive in the sense 
that we impose the geometry given in Eq. (\ref{geometry}) and then get 
the solution by satisfying the Einstein-Lanczos equations as well as 
the Klein-Gordon equations. In particular, using Eq.
 (\ref{geometry}), it is not difficult to show 
that Eqs. (\ref{EL1a}) and (\ref{EL2a}) imply, respectively,
\begin{eqnarray}
&&{\phi'}^2 + {\chi'}^2 = \frac{3 b^2}{\kappa} \frac{1}{(b^2 w^2 + 1)^3},
\label{kkex1}\\
&& W(\phi,\chi) = -\frac{ 3}{ 8 \kappa \epsilon} \frac{2 b^4 w^4+ 4 b^2 w^2-1}{(b^2 w^2+1)^2}.
\label{kkex2}
\end{eqnarray}
From Eq. (\ref{kkex1}) we deduce that $\phi$ and $\chi$ are given by 
\begin{eqnarray}
\phi(w) &=& \frac{v}{\sqrt{2}} \biggl(1 + \frac{b w}{\sqrt{b^2w^2 + 1}}\biggr)^{3/2}, 
\label{kink}\\
\chi(w) &=& \frac{v}{\sqrt{2}} \biggl(1 - \frac{b w}{\sqrt{b^2 w^2 + 1}}\biggr)^{3/2}.
\label{antikink}
\end{eqnarray}
provided the arbitrary constants $a_0$ and $v$ are such that
\begin{equation}
a_{0} = 2 \sqrt{\epsilon} \,b, \qquad v^2 = \frac{4}{3 \kappa}.
\label{parkak}
\end{equation}
The first of these two relations is necessary in order to write Eqs. (\ref{kkex1}) 
and (\ref{kkex2}) while the second relation is essential to solve 
Eq. (\ref{kkex1}) in terms of Eqs. (\ref{kink}) and (\ref{antikink}).
Knowing the form of the field profiles, 
the potential can be determined from Eq. (\ref{kkex2}) by adopting the following ansatz:
\begin{equation}
W(\phi,\chi) = {\mathcal A} (\phi^2 + \chi^2)^2 + {\mathcal B} (\phi^2 + \chi^2) + {\mathcal C}  + {\mathcal L}(\phi,\chi).
\label{ans}
\end{equation}
The functional ${\mathcal L}(\phi,\chi)$ vanishes exactly on the classical solution given by Eqs. (\ref{kink}) and (\ref{antikink}) but its derivatives do contribute to the Klein-Gordon equations. In fact Eq.
 (\ref{kgex}) can then be used to determine 
 ${\mathcal L}(\phi,\chi)$. Using Eqs. (\ref{kink}) and (\ref{antikink}) into Eq. (\ref{ans}) and recalling 
Eq. (\ref{kkex2}), 
the coefficients appearing in Eq. (\ref{ans}) are determined to be
\begin{equation}
{\mathcal A} = \frac{1}{8\kappa \epsilon v^4}= \frac{3}{32 v^2}
\qquad {\mathcal B} = - \frac{1}{\kappa \epsilon v^2} = 
- \frac{3}{4\epsilon},
\qquad {\mathcal C} = \frac{5}{4\kappa \epsilon} = \frac{15}{16} \biggl(\frac{v^2}{\epsilon}\biggr),
\label{coeff}
\end{equation}
where the second equality in each of the three relations follows 
by eliminating $\kappa$ according to Eq. (\ref{parkak}) (second
equality).
Inserting then Eqs. (\ref{geometry}), (\ref{ans}) and 
(\ref{kink})--(\ref{antikink}) into Eq. (\ref{kgex}), the 
functional form of ${\mathcal L}(\tilde{\phi},\tilde{\chi})$ can be determined:
\begin{equation}
{\mathcal L}(\tilde{\phi},\tilde{\chi}) = \frac{7}{2} \frac{v^2}{\epsilon}(|\tilde{\phi}|^{2/3} + 
|\tilde{\chi}|^{2/3} - 1) ( 1 - \tilde{\phi}^2 - \tilde{\chi}^2)^2,
\label{L}
\end{equation}
 where, for notational convenience, we defined the two rescaled fields 
$\tilde{\phi}= \phi/(2 v)$ and $\tilde{\chi} = \chi/(2 v)$. 
Using the second relation in Eq. (\ref{parkak}) into Eq. (\ref{coeff})
to eliminate $\kappa$ in favor of $v^2$, the complete form of the potential
becomes, in terms of $\tilde{\phi}$ and $\tilde{\chi}$, 
\begin{equation}
W(\tilde{\phi}, \tilde{\chi}) = \frac{3 v^2}{2\epsilon} (\tilde{\phi}^2 + 
\tilde{\chi}^2)^2 - \frac{3 v^2}{\epsilon} (\tilde{\phi}^2 + \tilde{\chi}^2) + 
\frac{15}{16} \frac{v^2}{\epsilon} + \frac{7}{2} \frac{v^2}{\epsilon} [ 
|\tilde{\phi}|^{2/3} +|\tilde{\chi}|^{2/3} -1] [ 1 - \tilde{\phi}^2 - \tilde{\chi}^2]^2.
\label{W}
\end{equation}
The solution given in Eqs. (\ref{kink}) and (\ref{antikink}) 
implies, necessarily, that $\phi >0$ and $\chi>0$. 
However, it appears from the analytical form of the 
potential that also $\phi \to -\phi$ or $\chi\to - \chi$ lead to acceptable 
solutions and this is the rationale for the absolute values in Eqs. (\ref{L}) 
and (\ref{W}).
\begin{figure}
\begin{center}
\begin{tabular}{|c|c|}
      \hline
      \hbox{\epsfxsize = 7.6 cm  \epsffile{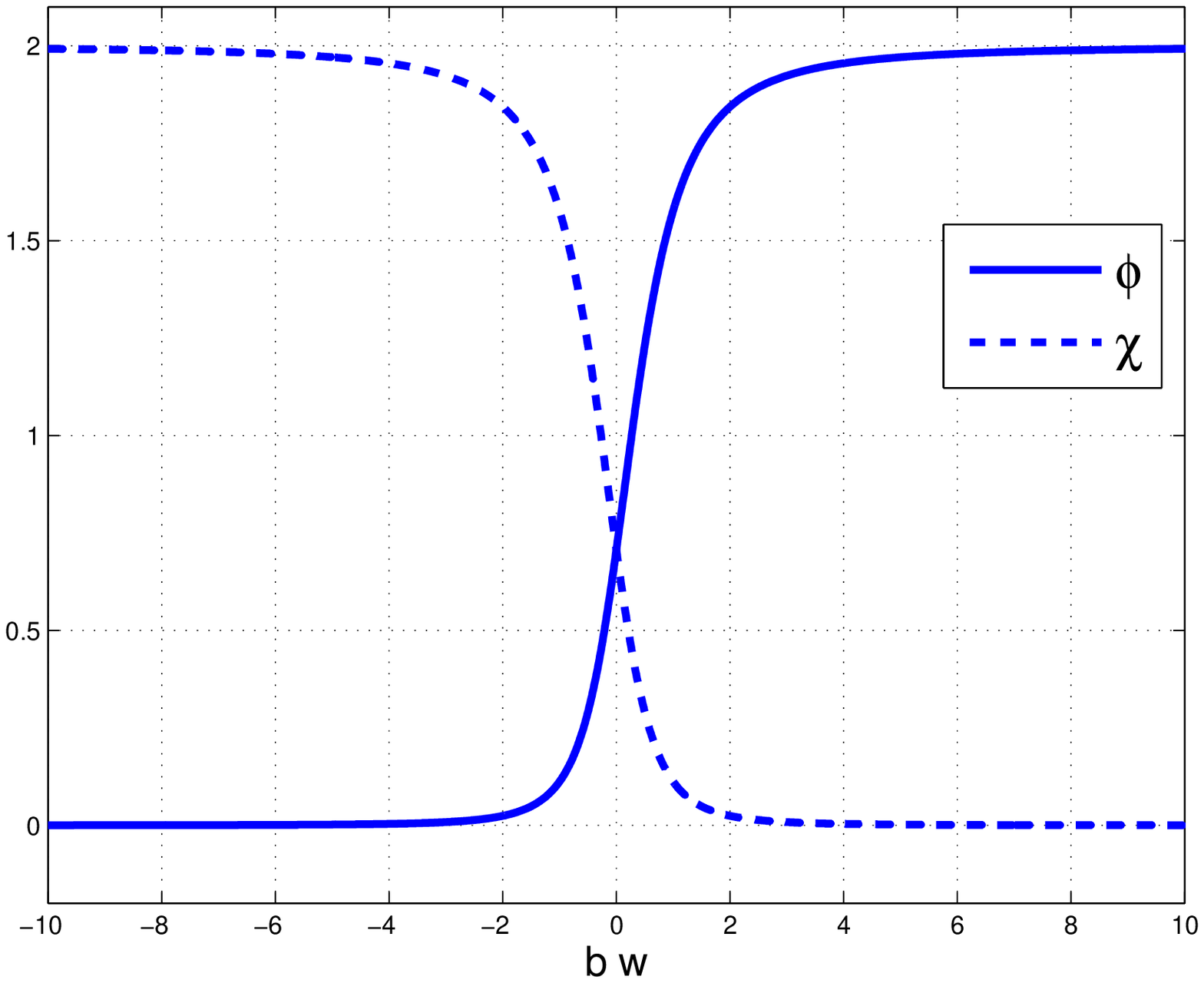}} &
      \hbox{\epsfxsize = 7.6 cm  \epsffile{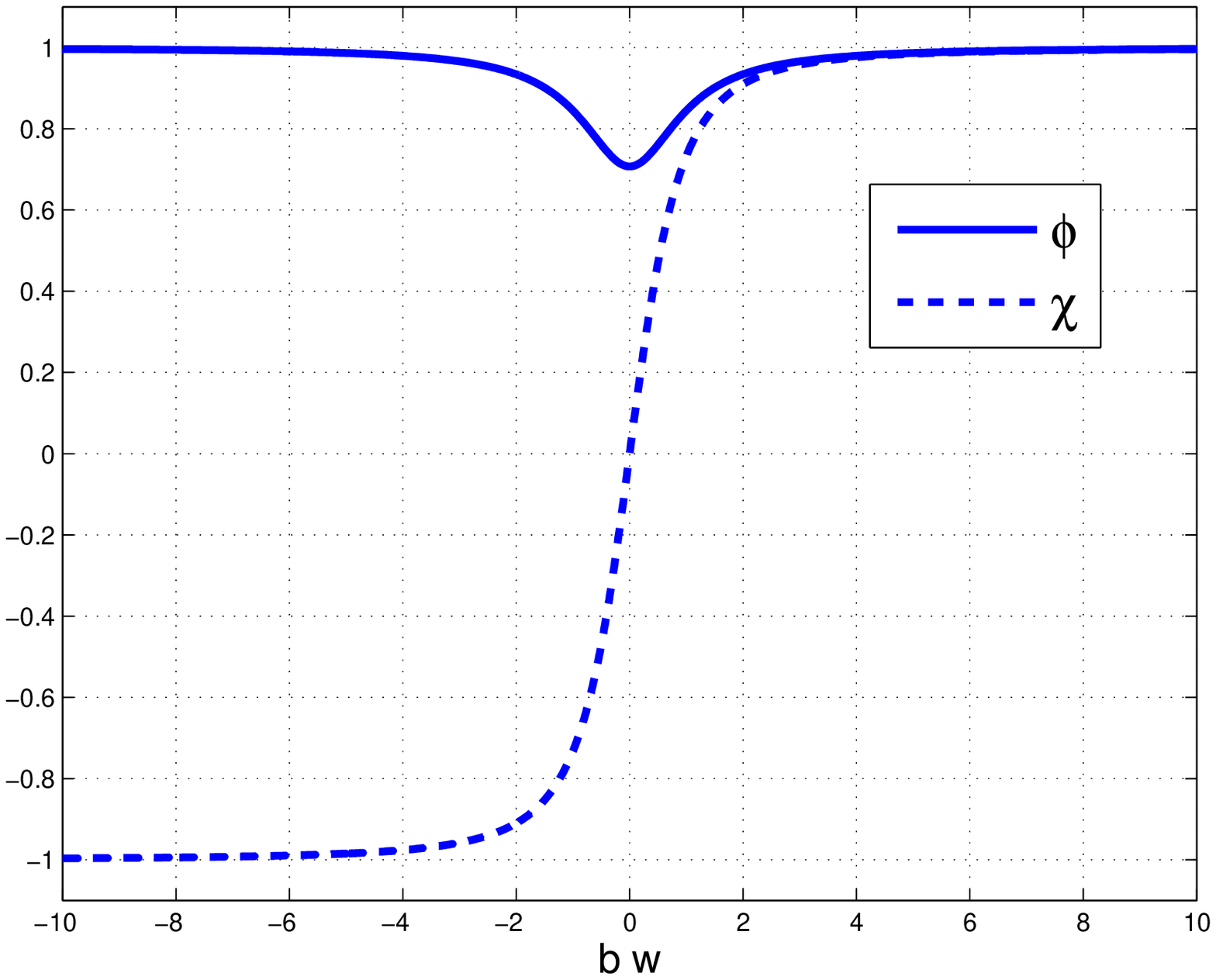}}\\
      \hline
\end{tabular}
\end{center}
\caption[a]{The kink-antikink solution (left plot) and the trapping-bag solution 
(right plot) are illustrated as a function of the bulk radius.}
\label{F1}
\end{figure}
In Fig. \ref{F1} (plot at the left hand side) 
the kink-antikink solution of Eqs. (\ref{kink})--(\ref{antikink}) is reported as a function of the rescaled bulk radius $b w$. 
In the case of one spatial dimension, spatial infinity consists of two points, i.e. 
$\pm \infty$; a topological charge is then customarily defined for the characterization  
of  $(1+1)$-dimensional defects such as the ones 
arising in the case of sine-Gordon system \cite{RAJA}. 
In the case of the kink-antikink system the topological charges 
can be defined as
\begin{equation}
Q_{\phi} = \frac{1}{2\pi} \int_{-\infty}^{\infty} \frac{\partial \phi}{\partial w} d w, \qquad Q_{\chi} = 
\frac{1}{2\pi} \int_{-\infty}^{\infty} \frac{\partial \chi}{\partial w} d w.
\label{TC}
\end{equation} 
Inserting the explicit solutions of Eqs. (\ref{kink}) and (\ref{antikink}) 
into Eq. (\ref{TC}) it is easy to find that $Q_{\phi} = - Q_{\chi} = v/\pi$.

By slightly modifying the form of the potential obtained in the case 
of the kink-antikink system one can obtain solutions of a different kind.
In $(1+1)$ dimensions these solutions are known as trapping 
bag solutions.
Exactly with the same procedure described above it can be shown that 
the following field profiles 
\begin{eqnarray}
&& \phi(w)= \frac{v}{2\sqrt{2}} \biggl[ \biggl( 1 + \frac{b w }{\sqrt{b^2 w^2+ 1}}\biggr)^{3/2} + \biggl( 1 - \frac{b w}{\sqrt{b^2 w^2+ 1}}\biggr)^{3/2} \biggl],
\label{phnt}\\
&& \chi(w) =  \frac{v}{2\sqrt{2}} \biggl[ \biggl( 1 + \frac{b w}{\sqrt{b^2 w^2+ 1}}\biggr)^{3/2} - \biggl( 1 - \frac{b w}{\sqrt{b^2 w^2 + 1}}\biggr)^{3/2} \biggl],
\label{chnt}
\end{eqnarray}
are solutions of the evolution equations previously deduced for the following 
 choice of the potential
\begin{eqnarray}
W(\phi,\chi) &=& \frac{3 v^2}{\epsilon} (\tilde{\phi}^2 + 
\tilde{\chi}^2)^2 - \frac{3 v^2}{\epsilon} (\tilde{\phi}^2 + \tilde{\chi}^2) + 
\frac{15 v^2 } {32 \epsilon} 
\nonumber\\
&+& 
\frac{7 v^2}{\epsilon} ( |\tilde{\phi} + \tilde{\chi}|^{2/3} + |\tilde{\phi} - \tilde{\chi}|^{2/3} -1)\biggl[ \frac{1}{2} - \tilde{\phi}^2 - \tilde{\chi}^2\biggr]^2.
\label{Wnt}
\end{eqnarray}
In this case the Einstein-Lanczos equations, i.e. Eqs. (\ref{EL1})--(\ref{EL2}), are satisfied only if 
\begin{equation}
 a_{0} = 2 \sqrt{\epsilon} b,\qquad v^2 = \frac{8}{3\kappa},
\label{parnt}
\end{equation}
i.e. the $a_0$ is the same as in Eq. (\ref{parkak}) while the relation of $v^2$ to $\kappa$ is different.
It is clear that Eqs. (\ref{phnt})--(\ref{chnt}) look like being the sum 
and the difference of the two profiles discussed above in  Eqs. 
(\ref{kink})--(\ref{antikink}). The system under consideration is, however,  intrinsically 
nonlinear and, therefore, 
 the compatibility of the solutions (\ref{phnt})--(\ref{chnt}) entails necessarily 
 a different relation between $v^2$ and $\kappa$ (compare Eqs. (\ref{parkak})
 and (\ref{parnt})) and also a slightly different form of the potential.

In Fig. \ref{F1} (plot at the right) the analytical solution of Eqs. (\ref{phnt}) 
and (\ref{chnt}) is illustrated for $v = 1$. From Fig. \ref{F1} it is 
also clear the rationale for the terminology employed in naming these 
solutions. The $\phi$ field is the "bag" that "traps" the $\chi$ field.
As already mentioned this type of trapping-bag solutions can be 
found, in $(1+1)$ dimensions, and with an appropriate 
nonlinear potential possessing a global $U(1)$ symmetry 
\cite{RAJA2,MONT,SOU}.
By inserting Eqs. (\ref{phnt}) and (\ref{chnt}) into Eq. (\ref{TC}), it is 
easy to show that while $Q_{\phi} =0$, $Q_{\chi} = v/\pi$. So, while 
the $\phi$ field illustrates a non-topological profile, the $\chi$ 
field is still topological.

The constructive technique exploited in the present paper can be extended in other cases when, for instance, the form of the 
underlying geometry is  different from the one of Eq. (\ref{geometry}).
In particular, it might be interesting to discuss the warp factor
\begin{equation}
a(w) = a_1 [(b w)^{2\nu} +1]^{-\frac{1}{2\nu}},
\label{gen}
\end{equation}
where $\nu\geq 1$ is an integer parameter. Notice that for $\nu =1$ 
Eq. (\ref{gen}) gives exactly Eq. (\ref{geometry}). For $\nu >1$ 
${\mathrm AdS_{5}}$ is always recovered, asymptotically, for 
large absolute value of the bulk radius.
Single field defects (both topological and non-topological) arising in the geometry (\ref{gen}) have been 
analyzed  in \cite{MGNT} in the case of Einstein-Hilbert gravity. It would 
be interesting to generalize these solutions to the case of Einstein-Lanczos gravity and in the presence of a pair of scalar degrees of freedom.
Along similar lines it seems also reasonable to think about the possibility 
of  multi-defects, i.e. gravtating profiles of two (or more) scalar degrees 
of freedom.

In the present investigation it has been argued that there may be a non-trivial 
interplay between five-dimensional 
Gauss-Bonnet gravity  and the
presence of unusual defects that may arise when two scalar 
degree of freedom are simultaneously present. 
Solutions describing both kink-antikink profiles and trapping bags have been presented.
While the possibility of qalitatively similar profiles in nonlinear $(1+1)$ dimensional 
field theories has been established in a number of different ways, 
in the context of five-dimensional Gauss-Bonnet gravity no attention 
has been payed to these configurations, to the best of our knowledge.
One of the interesting features of the obtained solutions is 
that the geometry which solves the Einstein-Lanczos equations 
in the presence either of kink-antikink profiles or in the 
presence of trapping bag profiles is always of ${\mathrm AdS}_{5}$.
More specifically the warp factor tends to  ${\mathrm AdS}_{5}$
for $|w|\to \infty$. Close to the core of the defect, i.e. 
for $|w| \to 0$ the geometry is always regular (i.e. all curvature invariants are regular); both $\phi$ and $\chi$ (as well as $\phi'$ and $\chi'$) are finite and regular.

\end{document}